\documentclass[conference]{IEEEtran}
\usepackage{graphicx} 
\usepackage{amssymb}
\usepackage{amsmath}
\usepackage{xcolor}
\usepackage{bm}
\usepackage{float}
\usepackage{makecell}
\usepackage{url}
\usepackage{hyperref}
\usepackage{threeparttable}
\usepackage{longtable}
\usepackage{titlesec}
\usepackage{nameref}
\usepackage{enumitem}
\usepackage{algorithm}
\usepackage{algpseudocode}
\usepackage{cite}
\usepackage{gensymb}
\usepackage{threeparttable}
\usepackage{amsthm}

\theoremstyle{remark}
\newtheorem{remark}{Remark}

\title{Energy Management for Solar-Powered Electric-Bus Charging Station: A Data-Driven Method\\
\small
\author{
 \IEEEauthorblockN{Xiaoting Wang, Supun Amarathunga, Pasan Gunawardena, Gregory Kish, Yunwei (Ryan) Li}
\IEEEauthorblockA{ 
\textit{Department of Electrical and Computer Engineering}, \textit{University of Alberta}, Edmonton, Canada \\
xiaotin5@ualberta.ca, amarathu@ualberta.ca, pasan@ualberta.ca, gkish@ualberta.ca, yunwei.li@ualberta.ca
}
    }
\normalsize
\thanks{This work is funded by the }
}
\date{October 2024}

\begin{document}

\maketitle

\begin{abstract}
 This paper presents a flexible energy management system (EMS) for an electric bus charging station (EBCS) that integrates renewable generation, energy storage, and electric bus (EB) charging while accounting for uncertainties in solar PV output, electricity prices, and EB arrival/departure state of charge. A data-driven polynomial chaos expansion surrogate is developed from a limited set of uncertainty samples, and a nonparametric inference method is used to enrich the input data when historical data is limited. Case studies on a solar-powered EBCS with 20 EBs demonstrate the effectiveness of the proposed EMS and data-driven method. 
\end{abstract}


\section{Introduction}
Electric buses (EBs) are increasingly being adopted by public transit agencies because of their environmental and economic advantages. In parallel, renewable-powered (e.g., solar PV) EB networks integrated with energy storage systems (ESS) are emerging as a promising approach to address growing energy and environmental challenges, including reducing carbon emissions, increasing savings, and improving the efficiency of power grid operations \cite{ren2022optimal}, \cite{pierrou2024optimal}. To fully realize these benefits, a flexible energy management system (EMS) is critical for EB charging stations (EBCSs), enabling optimal energy utilization while maintaining grid stability, operational efficiency, and economic performance. 

Several methods have been proposed for managing EBCSs with renewable generation, EBs, and ESS \cite{arif2021novel,Yu2019electric,manzolli2025electric,Zhuang2021,Wong2013}. For example, \cite{arif2021novel} develops an MILP-based EMS to maximize depot profit, 
but considers only a few deterministic daily scenarios without uncertainty analysis. To address solar PV uncertainty, Yu et al. \cite{Yu2019electric} propose a scenario-based chance-constrained model 
that jointly optimizes investment and operating costs, 
explicitly modeling PV and bus state of charge (SoC) uncertainties; however, it 
assumes fixed electricity prices, leaving the impact of price uncertainty unaddressed.
Another commonly used approach is Monte Carlo (MC) simulation \cite{Wong2013}, which is widely applied to investigate the impacts of uncertainty, yet its high computational burden limits its application. 

Recently, data-driven methods have been introduced to improve the computational efficiency of uncertainty analysis \cite{Jiang2023,Wang2018,chen2021data}. Jiang et al. \cite{Jiang2023}, for instance, propose a data-driven low-rank approach to assess the impact of uncertain EV loads on distribution system operation. Polynomial chaos expansion (PCE), another widely used surrogate modeling technique, has been combined with data-driven approaches and applied to power system studies such as probabilistic transfer capability \cite{wang2021} and security margin assessment \cite{Xu2020}, showing strong potential to reduce computational cost using only a small number of input samples. However, their effectiveness in modeling and propagating electricity price uncertainty within EMS for EB charging scheduling remains largely unexplored.

In this paper, we develop a flexible EMS for an EBCS that jointly coordinates renewable generation, ESS, and electric bus (EB) charging loads. The proposed framework explicitly accounts for uncertainties in solar PV generation, electricity prices, and EB arrival and departure SoC under a given parking schedule. To enable fast yet accurate optimal scheduling, we employ a data-driven polynomial chaos expansion (DDPCE) approach to construct a surrogate model for the EMS using only a limited number of uncertain input samples. In addition, a nonparametric inference method is used to infer and enrich the uncertain input samples when historical data are scarce. The effectiveness of the proposed approach is demonstrated through case studies based on real-world solar power and electricity price data for a solar-powered EBCS.

\section{The Proposed Energy Management System for Electric Bus Charging Station} \label{sec:ems}

This paper studies an EBCS equipped with solar PV, an ESS, grid interconnection, and EB loads. The goal is to produce a day-ahead charging schedule over each bus’s specified parking window while minimizing total operating cost, including penalties for any load shedding. The formulation explicitly models uncertainty in PV generation, electricity prices, and the buses’ arrival/departure SoC.


\subsection{Objective Function}

The objective of the proposed EBCS management framework is to minimize total operating cost under a day-ahead operating schedule. Specifically, the formulation incorporates the uncertainties from PV generation and EB demand alongside market electricity price trajectory:
%
\begin{equation}
    \mathrm{minimize} \quad  \lambda = \sum_{t\in\mathcal{T}}^{}[C_{\mathrm{im}}^t P_{\mathrm{im}}^t-C_{\mathrm{ex}}^tP_{\mathrm{ex}}^t +C_{\mathrm{shed}}^t P_{\mathrm{shed}}^t]\Delta t
    \label{eq:obj}
\end{equation}
\noindent where $\lambda$ denotes the minimum operating cost, which may be positive, zero, or negative; a negative value indicates a net benefit. $C_{\mathrm{im}}^t$ denotes the electricity buying price from the grid to customer, and $C_{\mathrm{ex}}^t$ denotes the electricity selling price from customer to the grid. $C_{\mathrm{shed}}^t$ denotes the penalty to shed load. $P^t_{\mathrm{im}}$ is power import from grid; $ P^t_{\mathrm{ex}} $ is the power export from the charging station to the grid; $P^t_{\mathrm{shed}}$  is the load shedding; $ \Delta t $ is the duration of the time interval. $\mathcal{T}=\{t_0,\cdots,t_{T}\}$ is the total time period index set.

\subsection{Power Balance}
The EMS coordinates grid exchange, solar generation, and ESS charge/discharge to supply the EBs at the EBCS. Accordingly, the following power balance constraints are imposed between generation and consumption in the EBCS:
\begin{equation}
P^t_{\mathrm{im}} + P^{t}_{\text{ESS,dis}} + P^t_{\text{PV}} + P^t_{\mathrm{shed}}= P^t_{\mathrm{ex}}  + P^t_{\text{ESS,ch}} + P^t_{\text{EB}}
\label{eq:power_balance}
\end{equation}
\noindent where $P^{t}_{\text{ESS,dis}}$ is the discharging power of the ESS; $P^t_{\text{ESS,ch}} $ is the charging power of the ESS; $P^t_{\text{PV}}$ is the solar generation power, and $P^t_{\text{EB}}$ is the charging load of EBs.

The power imported from or exported to the grid at time interval $t$ is constrained by the maximum allowable grid power.
\begin{equation}
0 \leq P^t_{\text{im}} \leq \bar{P}_{\mathrm{im}} u^t_{\text{G}}
\label{eq:gen_limits}
\end{equation}
\begin{equation}
0 \leq P^t_{\text{ex}} < \bar{P}_{\mathrm{ex}}  (1 - u^t_{\text{G}})
\label{eq:gen_limits2}
\end{equation}

\noindent where $\bar{P}_{\mathrm{im}}$ is the maximum amount of power that can be bought from the grid. $\bar{P}_{\mathrm{ex}}$ is the maximum amount of power that can be sold from the station to the grid. $u^t_{\text{G}}$ is a binary variable that decides purchasing/selling status of power from/to grid at time interval $t$.

\subsection{EB Load Modeling}
Transit services plan EB deployment by determining feasible routes and schedules and aggregating them into operational groups, referred to as EB blocks. Each block comprises multiple routes and driver shifts, usually concentrated in the morning or evening peak period. These assigned blocks enable realistic estimation of EB parking (layover/charging) times. For simplicity, this paper considers the EB charger operates in charge-only mode, i.e., no B2G power transfer. We consider $\mathcal{N}$ EBs at the transit network and treat the EB charging load as flexible load under given parking schedule.   For bus $n \in \{1,\ldots,\mathcal{N}\}$, define its daily parking (layover) schedule as the ordered set of arrival–departure pairs \cite{Amarathunga2025}:
\begin{equation}
  X_n \triangleq \big\{(t_{1,a}^n, t_{1,d}^n),\,(t_{2,a}^n, t_{2,d}^n),\,\ldots,\,(t_{K_n,a}^n, t_{K_n,d}^n)\big\}
  \label{eq:parking_time}
\end{equation}
where \(t_{k,a}^n\) and \(t_{k,d}^n\) denote, respectively, the \(k\)-th arrival at and departure from the EBCS within the day.

Let $P^{t}_{\mathrm{EB},n}$ denote the 
 charging power of $n$-th EB at $t$-th time interval. Then, the total EB load $P_{\mathrm{EB}}^t$ at time $t$ is the summation of each EB load:
 \begin{equation}
P^t_{\text{EB}} = \sum_{n \in \mathcal{N}} P^t_{\text{EB},n}
\label{eq:eb_load}
\end{equation}
The aggregate EB charging load at time $t$ is shaped by grid constraints, available solar power, the ESS state of charge SoC, and operational priorities such as imminent EB departures and the day–night period. Accordingly, the total EB load is regulated to lie within an allowable range:
\begin{equation}
  \underline{P}_{\mathrm{EB}} \;\le\; P_{\mathrm{EB}}^t \;\le\; \overline{P}_{\mathrm{EB}} ,
\end{equation}
where \(\underline{P}_{\mathrm{EB}}\) and \(\overline{P}_{\mathrm{EB}}\) denote the minimum and maximum admissible charging power, respectively. 
%
Assume each EB may charge whenever it returns to the EBCS. $P^{t}_{\mathrm{EB},n}$ should satisfy: 
\begin{equation}
P_{\text{EB},n}^{t} = 
\begin{cases} 
p^{t}_{n}, & t \in X_n \\ 
0, & t \in \mathcal{T} \setminus X_n 
\end{cases}
\label{eq_pev}
\end{equation}
where $p_{n}^{t}$ denotes the charging power of the $n$-th EB at $t\in X_{n}$:
\begin{equation}
0 < p_n^{t} < \bar{P}_{\mathrm{c}}
\label{eq:p_t}
\end{equation}
$\bar{P}_{\mathrm{c}}$ represents the maximum charging power of the EB chargers at the EBCS, assuming all chargers have identical rated capacities.
Furthermore, the SoC of the $n$-th EB at the end of interval $t$, denoted $\mathrm{SoC}_t^{n}$ (in \%), is updated from its previous value and the charging power as 
\begin{equation}
\text{SoC}_t^n = \text{SoC}_{t-1}^n + \frac{\eta_{\text{EB}} \cdot p_t^n \Delta t}{E_{\text{EB}}} \times 100\%\,\,; \quad t \in X_n
\label{eq:soc_new}
\end{equation}
$E_{\text{EB}}$ is the EB battery energy capacity;  $\eta_{\text{EB}}$ is the charging efficiency; $\Delta t$ is the interval length; The initialization $\mathrm{SoC}_{t-1}^{n}$ at an arrival instant follows $ \text{SoC}_{k,a}^{n} $ as previously defined.

\subsection{ESS Charging and Discharging Modeling}
The evolution of the ESS SoC in interval $t$, denoted $\text{SoC}^t_{\text{ESS}}$, is governed by the charging and discharging power as
\small
\begin{equation}
\text{SoC}^{t}_{\text{ESS}} = \text{SoC}^{t-1}_{\text{ESS}} + \frac{\eta_c \cdot P^{t}_{\text{ESS,ch}}}{E_{\text{ESS}}} \times 100\% - \frac{P^{t}_{\text{ESS,dis}}}{\eta_d \cdot E_{\text{ESS}}} \times 100\%
\label{eq:soc}
\end{equation}
\normalsize
\noindent where $\eta_c$ and $\eta_d$ are the ESS charging and discharging efficiencies, respectively, and $E_{\text{ESS}}$ is the ESS energy capacity. The ESS charging and discharging powers are limited by battery characteristics and by the capacity of the available chargers at the EBCSs:
\begin{equation}
0 < P^{t}_{\text{ESS,ch}} < \bar{P}_{\text{ESS}} U^t_{\text{ESS}}
\label{eq:ESS_charge}
\end{equation}
\begin{equation}
0 < P^{t}_{\text{ESS,dis}} < \bar{P}_{\text{ESS}} (1 - U^t_{\text{ESS}})
\label{eq:ESS_discharge}
\end{equation}
\noindent Herein, $\bar{P}_{\text{ESS}}$ is the maximum ESS charge/discharge power and $U^t_{\text{ESS}}$ is a binary variable,  indicating the ESS operating mode (charging when $U^t_{\text{ESS}}=1$, discharging when $U^t_{\text{ESS}}=0$). The ESS SoC is maintained within predefined limits to protect its state of health (SoH):
\begin{equation}
\underline{\text{SoC}}_{\text{ESS}} < \text{SoC}^t_{\text{ESS}} < \overline{\text{SoC}}_{\text{ESS}}
\end{equation}
\begin{equation}
\text{SoC}^{t_0}_{\text{ESS}} = \text{SoC}^{t_T}_{\text{ESS}}
\label{eq:soc_initial}
\end{equation}

\noindent Here, $\underline{\text{SoC}}_{\text{ESS}}$ and $\overline{\text{SoC}}_{\text{ESS}}$ denote the lower and upper SoC bounds, respectively. Constraint~\eqref{eq:soc_initial} enforces a daily cycle by setting the initial and terminal SoC to be equal.
In this paper, because solar generation, electricity prices, and EB loads fluctuate, and the SoC of EBs upon arrival is random, we treat all decision variables and objective values as random outputs.
For the simulation study in Section~\ref{sec:case_study}, we use measured solar power, electricity price, and EB arrival and departure SoC data from an operating transit system in Edmonton.
\section{The Proposed Data-Driven Framework for EMS}
\subsection{The moment-based DDPCE} \label{sec:pce_method}
Now let the EMS model \eqref{eq:obj}–\eqref{eq:soc_initial} be represented as a mapping
$y=F(\bm{\zeta})$,
where $\bm{\zeta}$  collects the $\mathcal{D}$-dimensional uncertain inputs (e.g.,  $\bm{\zeta}=\{ P_{\mathrm{PV}}^t$, $C_{\mathrm{im}}^t\} \in \mathbb{R}^{\mathcal{D}}$) and
$y$  denotes the model random outputs (e.g., decisions variables like  $P_{\mathrm{im}}^t$ and $P_{\mathrm{ex}}^t$, and objective value $\lambda$).
The proposed data-driven method adopts the data-driven-based polynomial chaos expansion (DDPCE) method to construct a sparse finite spectral model $\hat{y} = \hat{F}(\bm{\zeta})$, where only a limited number of input-response pairs of $\{\bm{\zeta}, y\}$ is required to build this model. The model can be represented as the weighted summation of a series of low order orthogonal polynomials $\Psi(\bm{\zeta})$  \cite{wang2021}:
\begin{equation}
\label{eq:PCE}
    \begin{aligned}
    y\approx \hat{y} = \hat{F}(\bm{\zeta}) = \sum_{\bm{\beta} \in \mathcal{A}} a_{\bm{\beta}} \Psi_{\bm{\beta}}(\bm{\zeta})
    \end{aligned}
\end{equation}
where $\bm{\beta}$ is the multi-index for orthogonal polynomial basis $\Psi(\bm{\zeta})$, $\mathcal{A}$ is the corresponding index set collecting all possible $\beta$. $a_{\bm{\beta}}$ are the unknown coefficients. Specially, multi-variate polynomial basis $\Psi(\bm{\zeta})$ is constructed as the tensor product of univariate polynomial basis $\phi_j({\zeta_j}), \{j=1,\cdots, \mathcal{D}\}$:
\begin{equation}
   \Psi(\bm{\zeta}) = \prod_{j=1}^{\mathcal{D}}  \phi^{(\beta_j)}_j(\zeta_j) = \prod_{j=1}^{\mathcal{D}}\sum_{l=0}^{\beta_j} \rho_{l}^{(\beta_j)} \zeta_j^{l}
   \label{eq:tensor}
\end{equation}
where $\beta_j=\{0,1,\cdots,H\}$ is the degree of univariate polynomial basis $\phi^{(\beta_j)}_j$ and $H$ is the PCE order. Specially, the $q$-norm truncation being applied to improve efficiency and the corresponding $\mathcal{A}=\{\bm{\beta}\in \mathbb{N}^{\mathcal{D}}: {\|\bm{\beta}\|}_q \leq H\}$ \cite{blatman2009chaos}. ${\|\bm{\beta}\|}_q \leq H$ and ${\|\beta\|}_q = \left(\sum_{j=1}^{\mathcal{D}}\beta_j^q\right)^{\frac{1}{q}}$.   $\rho_{l}^{(\beta_j)}$ are the coefficients of $\phi_j$ that can be obtained using the moment-based method  \cite{wang2021}:
\begin{small}
\begin{equation}
\label{eq:matrix_coefficients}
	\left[\begin{array}{cccc}
	\mu_{0,j} & \mu_{1,j} &\ldots & \mu_{\beta_j,j} \\
	\mu_{1,j} & \mu_{2,j} &\ldots & \mu_{\beta_j+1,j} \\
	\vdots & \vdots & \vdots & \vdots \\
	\mu_{\beta_j-1,j} & \mu_{l,j} & \ldots & \mu_{2\beta_j-1,j} \\
	0 & 0 & \ldots & 1
	\end{array}\right] \left[\begin{array}{c}
	p_{0}^{(\beta_j)} \\ p_{1}^{(\beta_j)} \\\vdots \\ p_{{\beta_{j}}-1}^{(\beta_j)} \\ p_{\beta_j}^{(\beta_j)}
	\end{array}\right] = \left[\begin{array}{c}
	0\\0\\ \vdots \\ 0\\1
\end{array}\right]
\end{equation}
\end{small}

\noindent where $\mu_{s,j}$ denotes the $s$-th raw moment of $\zeta_j$, with $s=\{0,\cdots,2\beta_j-1\}$, $j=\{1,\cdots,\mathcal{D}\}$,  $\beta_j=\{0,1,\cdots,H\}$. $\mu_{s,j}$ can be computed directly from the the available sample data or from the probability distribution of $\zeta_j$.

Once $\Psi(\bm{\zeta})$ is constructed, unknown coefficients $a_{\bm{\beta}}$ are calculated by be minimizing the sum of the squared
residuals: 
\begin{equation}
    \sum_{m =1}^{\mathcal{M}_{\mathrm{tr}}}\left[ y^{(m)} - \sum_{\bm{\beta}\in{\mathcal{A}}}a_{\bm{\beta}} \Psi_{\bm{\beta}}(\bm{\zeta}^{(m)})\right]^2 
    \label{eq:min}
\end{equation}

\noindent where $\mathcal{M}_{\mathrm{tr}}$ denotes the size of the training dataset used to construct model~\eqref{eq:PCE}. We employ a greedy algorithm, orthogonal matching pursuit (OMP)~\cite{Marelli2018a} which, at each iteration, selects the basis $\Psi_{\boldsymbol{\beta}}$ most correlated with the current residual, appends it to the active set, and updates the coefficients $a_{\boldsymbol{\beta}}$ by the ordinary least-squares method on the active basis.

Once the multivariate polynomial basis $\Psi_{\boldsymbol{\beta}}(\boldsymbol{\zeta})$ and the coefficients $a_{\boldsymbol{\beta}}$ are determined, the moment-based DDPCE model~\eqref{eq:PCE} is fully specified. For a new dataset, one can evaluate the surrogate by feeding in new samples of $\boldsymbol{\zeta}$ to obtain approximations to the EMS solutions~\eqref{eq:obj}–\eqref{eq:soc_initial}, e.g., the minimum operating cost $\lambda$. The next subsection will present the implementation in detail.

\subsection{The Proposed PCE–based EMS}
In this subsection, we provide a detailed description of the proposed method for EMS of EBCS outlined step by step.

\noindent \textbf{Step 1 Data and Configurations.} Input $\bm{\zeta}_{\mathrm{tr}} = [\bm{P}_{\mathrm{PV,tr}}^{t},\bm{C}_{\mathrm{im,tr}}^t] \in \mathbb{R}^{\mathcal{M}_{\mathrm{tr}}\times \mathcal{D}}$ including the yearly day-ahead electricity price, and historical solar power. Provide the bus parking schedule. Preprocess EB SoC records by pairing arrival and departure SoC and aligning them with the corresponding parking windows.  Specify network and physical limits (e.g., charger ratings, ESS bounds, line limits). Forward $\boldsymbol{\zeta}_{\mathrm{tr}}$ to \textbf{Step 2}.

\noindent \textbf{Step 2 EMS Framework Construction.} Build the proposed EMS framework~\eqref{eq:obj}–\eqref{eq:soc_initial} and apply $\boldsymbol{\zeta}_{\mathrm{tr}}$ to the built framework to obtain the day-ahead charging schedule, the minimum cost $\lambda$, optimal grid exchange power ($P_{\mathrm{im}}^{t}, P_{\mathrm{ex}}^{t}$), ESS power output ($P_{\mathrm{ESS,ch}}^{t}$, $P_{\mathrm{ESS,dis}}^{t}$), EB loads $P_{\mathrm{EB},n}^{t}$, load shedding $P_{\mathrm{shed}}^{t}$, SoC variation of EB batteries.  

\noindent \textbf{Step 3 Generate Yearly Results and dataset.}  Generate the day-ahead schedule  results report obtained from \textbf{Step 2}. Let the response $y$ be the minimum cost $\lambda$. Pass input-response pairs $[\boldsymbol{\zeta}_{\mathrm{tr}},\boldsymbol{y}_{\mathrm{tr}}]$ to \textbf{Step 4}. 

\noindent \textbf{Step 4 Build Surrogate Model.}  Apply the moment-based PCE method presented in Section \ref{sec:pce_method} to build the PCE surrogate model \eqref{eq:PCE} for the EMS \eqref{eq:obj}–\eqref{eq:soc_initial} using \eqref{eq:tensor}-\eqref{eq:min}.

\noindent \textbf{Step 5 Scenario generation and evaluation.}
\begin{itemize}
    \item a) Apply the nonparametric inference  in Section~\ref{sec:data_infer} to estimate the distribution of solar PV power $P_{\mathrm{PV}}^{t}$.
    \item b) Generate additional $\mathcal{M}_{\mathrm{val}}$  sample points by sampling solar power $\bm{P}_{\mathrm{PV,val}}^{t}$ from the inferred distribution.
    \item c) Generate  additional electricity prices sample points $\bm{C}_{\mathrm{im,val}}^{t}$ based on a certain variation (e.g., 10\%) of the historical data.
    \item d) Construct the input matrix $\boldsymbol{\zeta}_{\mathrm{val}}=[\,\bm{P}_{\mathrm{PV,val}}^{t},\,\bm{C}_{\mathrm{im,val}}^{t}\,]\in\mathbb{R}^{\mathcal{M}_{\mathrm{val}}\times \mathcal{D}}$, evaluate the built surrogate~\eqref{eq:PCE} for each scenario, and collect the resulting minimum-cost values as $\boldsymbol{\lambda}_{\mathrm{val}}=[\,\lambda^{(1)},\ldots,\lambda^{(\mathcal{M}_{\mathrm{val}})}\,]$.
\end{itemize}

\noindent \textbf{Step 6 Surrogate Model Results Reports.} Compute the Statistics of $\lambda$, summarize the results reports.


\begin{remark}
  Note that in \textbf{Step 1}, inputs data are taken directly from real-world data. In Step 4, if the accuracy of the proposed method is not reached, go to \textbf{Step 5}. Otherwise, enrich the training sample data refer to methods described in \textbf{Step 5} a)-c) and go to \textbf{Step 2}.
\end{remark}

\begin{remark}
More generally, the constructed surrogate model \eqref{eq:PCE} can be used to approximate any EMS output $y$ (e.g., grid power exchange) that possesses a finite second-order moment.
\end{remark}
\subsection{Data Inference} \label{sec:data_infer}
The proposed EMS framework can be implemented directly with a modest amount of historical data. However, accurately characterizing the full distribution of EMS outcomes typically requires a large number of scenarios. To further assess the framework’s performance, additional data are therefore desirable. When historical data are limited, rather than imposing a specific parametric form (e.g., a Beta distribution for solar power as is common in the literature), we infer distribution of inputs $\bm{\zeta}$ (e.g., solar power $P_{\mathrm{PV}}^t$) using a non-parametric kernel density estimator (KDE), which allows us to generate more data with a parameter free model and can be expressed as \cite{Xu2020,silverman2018density}:
%
\begin{equation}
    \hat{f}_{\zeta}(\zeta) = \frac{1}{M\omega}\sum_{m=1}^{M}K \!\left(\frac{\|\zeta-\hat{\zeta}\|}{\omega}\right)
\end{equation}
where $M$ is the number of observations, $\omega >0$ is the bandwidth that controls smoothness. $K(\cdot)$,  
a nonnegative kernel function, is chosen as the 
standard Gaussian kernel. A common bandwidth choice is: $\omega = \left(\frac{4\hat{\sigma}^5}{3M}\right)^{1/5}
      \approx 1.06\,\hat{\sigma} M^{-1/5}$, with $\hat{\sigma}$ denoting the sample standard deviation.

\section{Simulation Studies} \label{sec:case_study}
To evaluate the performance of the proposed data-driven-based EMS, 
this paper considers a practical case study of EB charging scheduling.  In Section~\ref{sec:practical_case}, we first apply the EMS 
described in Section~\ref{sec:ems} using yearly real-world solar PV generation and electricity price data, as well as measured arrival and departure SoCs of the EBs. 
Given a specified parking schedule, the EB charging load is obtained from the optimal charging schedules produced by the EMS. In Section~\ref{sec:results}, we then develop 
the proposed data-driven EMS based on the results from Section~\ref{sec:practical_case} and evaluate its performance using additional sampled realizations of solar PV power 
and electricity prices, with MC simulations used as a benchmark.


In the case study, an EBCS with $1000~\text{m}^2$ of solar panels (15\% efficiency) is considered. 
The grid power import limit $\bar{P}_{\mathrm{im}}$ and export limit $\bar{P}_{\mathrm{ex}}$ are set to 
$500~\text{kW}$ and $100~\text{kW}$, respectively. The load shedding cost $C_t^{\mathrm{shed}}$ 
is set to $100~\text{cents/kWh}$. The maximum charging/discharging power of the ESS, 
$\bar{P}_{\text{ESS}}$, is $120~\text{kW}$. The minimum and maximum SoC levels of the ESS, 
$\underline{\text{SoC}}_{\text{ESS}}$ and $\overline{\text{SoC}}_{\text{ESS}}$, are maintained 
within 30\% and 90\%, respectively. The initial SoC of the ESS is assumed to be 
$330~\text{kWh}$ out of a total ESS energy capacity of $600~\text{kWh}$. Furthermore, 
a fleet of $20$ EBs is considered, and the EBCS is equipped with chargers with 
rated power of $60~\text{kW}$.
The EMS framework was implemented in a Python Jupyter notebook using Pyomo, while the proposed data-driven method was implemented in MATLAB R2024b.

\subsection{ A Practical Implementation of EMS for the EBCS} \label{sec:practical_case}
This case study applies real-world solar PV generation and electricity price data to the proposed EMS formulation in~\eqref{eq:obj}–\eqref{eq:soc_initial}. Specifically, the electricity buying price $C_{\mathrm{im}}^t$ and selling price $C_{\mathrm{ex}}^t$ are assumed to be identical and are taken from the 2023 hourly market pool prices published by the Alberta Electric System Operator (AESO)~\cite{AESO_ETS}. The real solar PV power data used in this section are obtained from  \color{black}National Solar Radiation Database (NSRDB)~\cite{NSRDB_Data}. \color{black}
In total, 365 scenarios are considered in this case. 
A 15-minute time interval is adopted, leading to $|\mathcal{T}| = 96$ time steps for each day-ahead schedule. The EB parking schedule is given in Table~\ref{table:parking_schedule}. Each EB is assumed to park at the EBCS twice per day ($k=2$) after completing its assigned service blocks.
\vspace{-8pt}
\begin{table}[htbp]
\caption{Parking Schedule of EBs}
\begin{center}
\begin{tabular}{c|c|c}
\hline
\textbf{Bus Name} & \textbf{\textit{Parking duration 1}}& \textbf{\textit{Parking duration 2}} \\
\hline
A,G,M,S & 00:00-06:00 & 14:00-18:00 \\
\hline
B,H,N,T & 06:00-14:00 & 18:00-24:00 \\
\hline
C,I,O & 09:00-13:00 & 21:00-03:00 \\
\hline
D,J,P & 03:00-09:00 & 14:00-21:00 \\
\hline
E,K,Q & 06:00-12:00 & 20:00-24:00 \\
\hline
F,L,R & 00:00-16:00 & 12:00-20:00 \\
\hline
\end{tabular}
\end{center}
\label{table:parking_schedule}
\end{table}
\vspace{-8pt}

\begin{figure}[htbp]
\centerline{\includegraphics[width=\columnwidth, height=5cm]{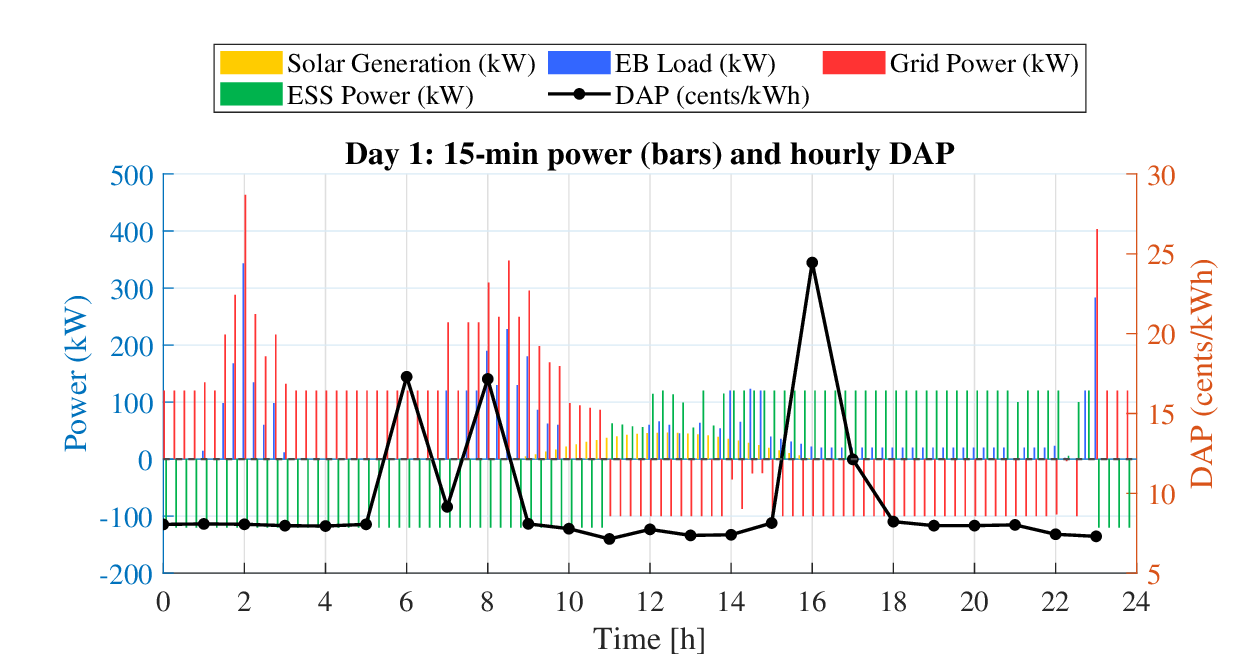}}
\caption{Hourly DAP and power exchange, ESS charge/discharge power, and EB total loads in 15-minutes interval for day 1.} 
\label{fig:power}
\end{figure}

\begin{figure}[htbp]
\setlength{\abovecaptionskip}{-4pt}
\setlength{\belowcaptionskip}{0pt}
\centerline{\includegraphics[width=0.4\textwidth]{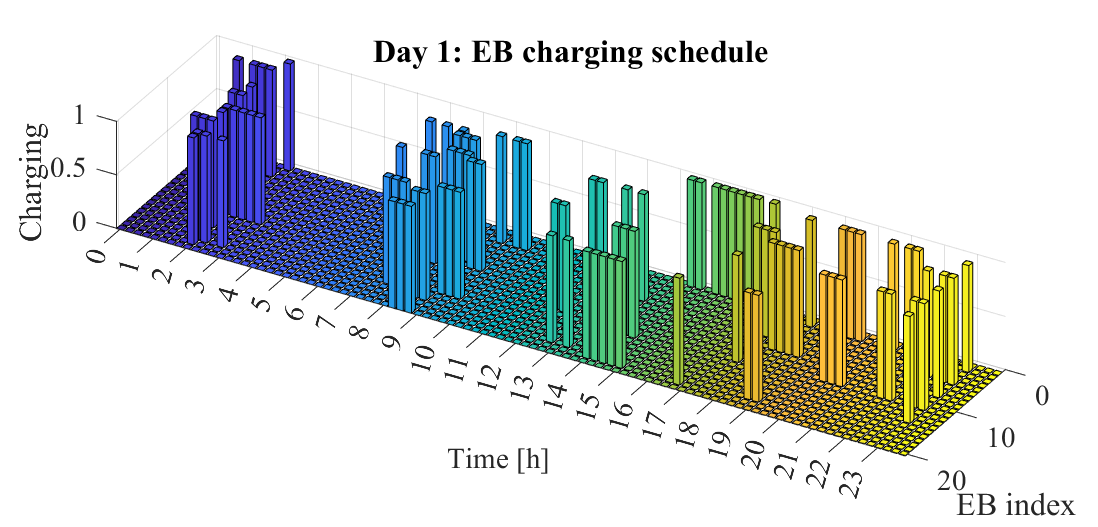}}
\caption{EB load charging state for day 1.}    
\label{fig:EB_load}
\vspace{-0.2in}
\end{figure}

\begin{figure}[htbp]
\setlength{\abovecaptionskip}{-4pt}
\setlength{\belowcaptionskip}{0pt}
\centerline{\includegraphics[width=0.38\textwidth]{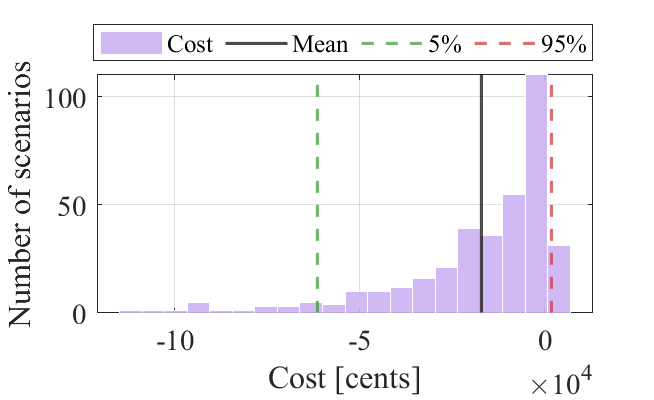}}
\caption{The histogram and statistics of minimum cost $\lambda$ for 365 scenarios.}  
\label{fig:cost}
\vspace{-0.22in}
\end{figure}

Once the input data and system configurations are prepared (Step 1), the EMS is solved (Step 2) and the results are summarized (Step 3). For example, Fig.~\ref{fig:power} presents the hourly DAP and the corresponding power exchange with the grid, ESS charging/discharging power, and EB total load at a 15-minute resolution for January 1. As expected, the ESS charges when the electricity price is low and discharges when the price is high and the EB load is relatively low. Fig.~\ref{fig:EB_load} shows the corresponding EB charging and discharging status. The minimum operating cost on the winter day is CAD~40.29, whereas on a
sunny summer day with high solar output (e.g., July~30), the operating
cost drops to CAD~$-38.60$, corresponding to a net profit of CAD~38.60. 
The probability distribution of the minimum daily cost is shown in Fig.~\ref{fig:cost}. Over the full year, the cost has a mean of CAD $–170.92$, with the 5th and 95th percentiles at CAD $–615.18$ and CAD $18.86$, respectively, highlighting the impact of electricity price and solar generation variability on operating cost. 
  
\vspace{-8pt}

\subsection{Results for the Proposed Data-Driven Method} \label{sec:results}
\begin{figure}[htbp]
\vspace{-0.22in}
\setlength{\abovecaptionskip}{-4pt}
\setlength{\belowcaptionskip}{0pt}
\centerline{\includegraphics[width=0.38\textwidth]{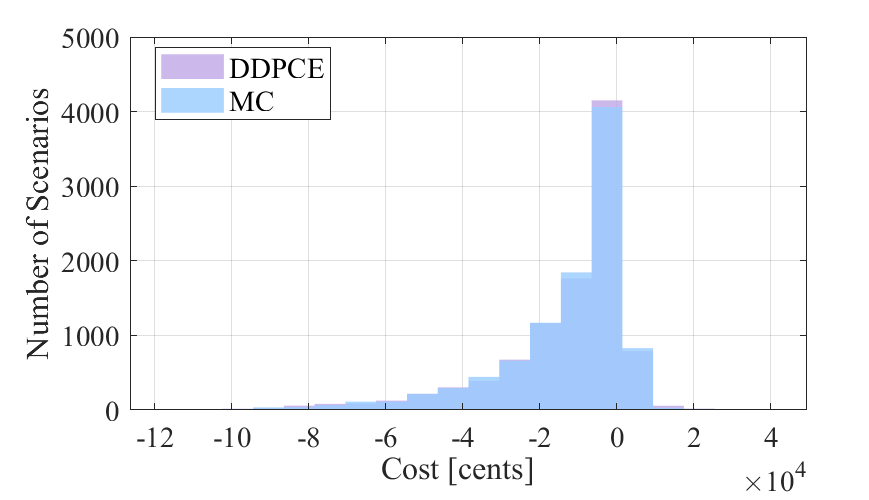}}
\caption{The comparison of histogram  of minimum cost $\lambda$ from DDPCE and the MC method. $\mathcal{M}_{\mathrm{val}} = 10000$ scenarios.}  
\label{fig:cost_ddpce}
\vspace{-0.2in}
\end{figure}

Based on the results in Section~\ref{sec:practical_case}, we enriched the input–response
pairs $[\boldsymbol{\zeta}_{\mathrm{tr}},\boldsymbol{y}_{\mathrm{tr}}]$ and passed the dataset to Step~4 to construct the
DDPCE model~\eqref{eq:PCE} for the EMS problem~\eqref{eq:obj}–\eqref{eq:soc_initial}, with the goal
of approximating the minimum cost (i.e., $\boldsymbol{y}_{\mathrm{tr}} = \boldsymbol{\lambda}$).
Specifically,
$\boldsymbol{\zeta}_{\mathrm{tr}} = [\boldsymbol{P}_{\mathrm{PV,tr}}^{t}, \boldsymbol{C}_{\mathrm{im,tr}}^{t}]
\in \mathbb{R}^{1200 \times 120}$, corresponding to $\mathcal{M}_{\mathrm{tr}} = 1200$ and
$\mathcal{D} = 120$. The PCE order is set to $H = 3$. In Step~5, $10{,}000$ validation samples
($\mathcal{M}_{\mathrm{val}}$) are generated to estimate the probabilistic characteristics of the cost
$\lambda$ using the constructed model. Fig.~\ref{fig:cost_ddpce} compares the proposed DDPCE
results with those obtained from the benchmark MC method, showing that the two curves almost
overlapping. The average cost estimated by DDPCE is CAD~$-127.60$, and the normalized error \% of this
average with respect to the MC estimate is $-1.94\%$, demonstrating the high accuracy of the
proposed approach.

\section{Conclusions}
This paper developed a flexible EMS for an EBCS to integrate renewable generation, ESS, and EB charging demand while explicitly accounting for uncertainties in solar power, electricity prices, and EB arrival/departure SoC. A DDPCE surrogate and nonparametric inference were employed to enable fast probabilistic scheduling from limited data, and case studies on a solar-powered EBCS with 20 EBs showed that the method effectively exploits PV and price variability and accurately estimates cost distributions with far fewer EMS evaluations than MC simulations.

\section*{Acknowledgment}
The authors would like to thank Professor Mahdi Shahbakhti and Dr.~Bahram Bahri of the Energy Mechatronics Laboratory (EML), Department of Mechanical Engineering, University of Alberta, for their valuable assistance with EB data collection.

\bibliographystyle{IEEEtran}
\bibliography{ems.bib}

\end{document}